\documentclass[aps,prb,preprint,groupedaddress,showpacs,amsmath,amssymb]{revtex4}

\usepackage{graphicx}% Include figure files
\usepackage{dcolumn}% Align table columns on decimal point
\usepackage{bm}% bold math

\begin{document}

\title{Heat capacity jump at $T_c$  and pressure derivatives of superconducting transition temperature in the Ba$_{1-x}$K$_x$Fe$_2$As$_2$ ($0.2 \leq x \leq 1.0$) series}

\author{Sergey L. Bud'ko$^{1}$, Mihai Sturza$^{2}$, Duck Young Chung$^{2}$, Mercouri G. Kanatzidis$^{2,3}$, and Paul C. Canfield$^{1}$}
\affiliation{$^{1}$Ames Laboratory, US DOE and Department of Physics and Astronomy, Iowa State University, Ames, IA 50011, USA}
\affiliation{$^{2}$Materials Science Division, Argonne National Laboratory, Argonne, Illinois 60439-4845, USA}
\affiliation{$^{3}$Department of Chemistry, Northwestern University, Evanston, Illinois 60208-3113, USA}

\date{\today}

\begin{abstract}
We present the evolution of the initial (up to $\sim 10$ kbar) hydrostatic, pressure dependencies of $T_c$ and of the ambient pressure jump in the heat capacity associated with the superconducting transition as a function of K - doping in the  Ba$_{1-x}$K$_x$Fe$_2$As$_2$ family of iron-based superconductors. The pressure derivatives show weak but distinct anomaly near $x \sim 0.7$. In the same concentration region $\Delta C_p|_{T_c}$ deviates from the $\Delta C_p \propto T^3$ scaling found for most BaFe$_2$As$_2$ - based superconductors. These results are consistent with a Lifshitz transition, and possible significant modification of the superconducting state, occuring near $x \sim 0.7$.

\end{abstract}

%\keywords{}

\maketitle

\section{Introduction}

For the last half decade a significant experimental and theoretical effort has concentrated on studies of the physical properties of Fe - based superconductors and related materials. \cite{joh10a,ste11a,joh11a} Of the several families of Fe-based superconductors discovered to date, the 122, {\it AE}Fe$_2$As$_2$ ({\it AE} = alkaline earth and Eu) family, is the most studied one. \cite{can10a,nin11a,man10a} This family offers possibilities for substitutions on all three crystallographic sites that can result in a complex combination of steric and carrier-doping effects while maitaining simplicity of the crystal structure. However, the main body of the published work on the 122 family has focused on the  {\it AE}(Fe$_{1-x}${\it TM}$_x$)$_2$As$_2$  series, in which the $3d-$, $4d-$, and sometimes $5d-$ transition metals substitute for Fe. This is due to the reasonable ease of growing large, high quality, homogeneous single crystals. 

Despite the fact that in the 122 system, superconductivity, with a $T_c \approx 38$ K,  was first reported on K-doping of BaFe$_2$As$_2$ \cite{rot08a}  and this value of $T_c$ is still holding a record for bulk superconductivity in the {\it AE}Fe$_2$As$_2$ - related materials, detailed studies of the complete Ba$_{1-x}$K$_x$Fe$_2$As$_2$ solid solution series are not so common, \cite{rot08b, che09a,avc12a} and concentrate mainly on the evolution of the crystallographic properties and the $x - T$ phase diagram. One of the reasons for this is a persistent difficulty in determination of the reliable procedure for synthesis of homogeneous  Ba$_{1-x}$K$_x$Fe$_2$As$_2$ single crystals in a wide range of K - concentration.

In the hole-doped, Ba$_{1-x}$K$_x$Fe$_2$As$_2$, series superconductivity is observed over a wide range of K - concentrations, $0.15 \lesssim x \leq 1.0$ \cite{rot08b, che09a,avc12a}  (as compared to $0.03 \lesssim x \lesssim 0.15$ for the electron-doped Ba(Fe$_{1-x}$Co$_x$)$_2$As$_2$ \cite{nin08a}).  For  underdoped Ba$_{1-x}$K$_x$Fe$_2$As$_2$ superconductivity and magnetism microscopically co-exist \cite{wie11a} (similarly to what was observed in the electron-doped Ba(Fe$_{1-x}${\it TM}$_x$)$_2$As$_2$ \cite{can10a,pra09a}). Near optimal doping, $x \sim 0.4$, several experiments and theoretical calculations give evidence for a nodeless, near constant, $s_\pm$ superconducting gap that changes sign between hole and electron pockets. \cite{din09a,chr08a,luo09a,gra09a,maz09a} Recent ARPES measurements \cite{nak11a} suggest that the  $s_\pm$ superconducting gap exists in a wide doping range, $0.25 \leq x \leq 0.7$.

KFe$_2$As$_2$ stands out among the members of the Ba$_{1-x}$K$_x$Fe$_2$As$_2$ series. The reported Fermi surface of KFe$_2$As$_2$ differs from that of the optimally doped  Ba$_{1-x}$K$_x$Fe$_2$As$_2$ having three hole pockets, two centered at the $\Gamma$ point in the Brillouin zone, and one around the $M$ point \cite{sat09a} with no electron pockets. Quantum criticality, and nodal or $d$-wave superconductivity in KFe$_2$As$_2$ was suggested in a number of publications. \cite{has10a,don10a,ter10a,don10b,rei12a,mai12a,tho11a}  Possible evolution from $s_\pm$ to $d$-wave in the  Ba$_{1-x}$K$_x$Fe$_2$As$_2$ series has also been discussed. \cite{rei12b,hir12a}

Given the debates present in the literature regarding the evolution of physical properties in the Ba$_{1-x}$K$_x$Fe$_2$As$_2$ series and its apparent difference from the extensively studied {\it AE}(Fe$_{1-x}${\it TM}$_x$)$_2$As$_2$ series, it is of importance to have a broad set of data on Ba$_{1-x}$K$_x$Fe$_2$As$_2$, in particular, data related to the superconducting state. In this work we present two datasets obtained for the Ba$_{1-x}$K$_x$Fe$_2$As$_2$ series with K-concentrations covering underdoped, optimally doped, and overdoped regions, up to the end compound, pure KFe$_2$As$_2$. The first set is comprised of the initial ($P \lesssim 10$ kbar) pressure dependencies of the superconducting transition temperatures, $T_c(P)$ and is alike the data reported for Ba(Fe$_{1-x}$Co$_x$)$_2$As$_2$. \cite{ahi09a} Such data have the potential to assess the possibility of equivalence of pressure and doping that was suggested for several 122 series. \cite{kim09a,dro10a,kli10a,kim11a} Moreover, under favorable circumstances such dataset can shed light on the details of superconductivity mechanism in the particular series. \cite{xio92a,cao95a,fie96a} The second set consists of the data on the evolution of the jump in heat capacity at the superconducting transition. Many  Fe - based, 122 superconductors follow the trend suggested in Ref. \onlinecite{bud09b} and expanded in Ref. \onlinecite{kim11b}, the so-called BNC scaling, $\Delta C_p|_{T_c} \propto T_c^3$. The underdoped and optimally doped  Ba$_{1-x}$K$_x$Fe$_2$As$_2$ appear to follow this trend as well, \cite{bud09b,kan10a} whereas the stoichiometric end-compound,  KFe$_2$As$_2$, clearly deviates from the trend. \cite{bud12a} Detailed study of the $\Delta C_p|_{T_c}$  in the whole range of K concentrations might clarify the evolution of superconducting properties in the series.

\section{Experimental details}

Homogeneous, single phase, Ba$_{1-x}$K$_x$Fe$_2$As$_2$ polycrystalline powders with $x$ = 0.2, 0.3, 0.4, 0.6, 0.7, 0.8, 0.9 and 1.0 were synthesized by a procedure similar to that reported earlier.\cite{avc12a} The starting reagents, KAs, BaAs, and Fe$_2$As were prepared by heating elemental mixtures at $450\,^{\circ}\mathrm{C}$, $650\,^{\circ}\mathrm{C}$, and $850\,^{\circ}\mathrm{C}$, respectively. For each composition a stoichiometric mixture of these binary compounds  was thoroughly ground to a uniform and homogeneous powder, loaded in an alumina crucible subsequently enclosed in a niobium tube and then sealed in a quartz tube, and pre-heated at $600\,^{\circ}\mathrm{C}$ for 12 h. This sintered mixture was then re-ground and pressed into a pellet which was again loaded in a niobium tube and sealed in a quartz tube. The pellet was heated to $800\,^{\circ}\mathrm{C} \leq T \leq1100\,^{\circ}\mathrm{C}$, depending on the composition, for 24 to 48 h followed by quenching to room temperature in air. The homogeneity of Ba$_{1-x}$K$_x$Fe$_2$As$_2$ polycrystalline powder was ensured by repeating this process multiple times. The structure and quality of the final crystalline powders of Ba$_{1-x}$K$_x$Fe$_2$As$_2$ were confirmed by x-ray powder diffraction and magnetization measurements and the $x$ values of all products were estimated by comparing the superconducting $T_c$'s observed with those precisely determined in the phase diagram reported previously. \cite{avc12a}  For the end-member sample with $x = 1.0$, the single phase product was prepared by a flux reaction of Fe/KAs at 1:6 ratio at $1000\,^{\circ}\mathrm{C}$ for 6 h. The pure crystals of KFe$_2$As$_2$ were then isolated by dissolving the excess KAs flux in alcohol under a nitrogen atmosphere.

The synthesis of Ba$_{1-x}$K$_x$Fe$_2$As$_2$ has been known to be delicate mainly because of high vapor pressures of K/As and an unfavorable kinetics toward the composition of optimally doped state $x = 0.4$ that exhibits the highest $T_c$ in the Ba$_{1-x}$K$_x$Fe$_2$As$_2$ series. Particularly for the underdoped region between $x = 0.3$ and $x = 0.15$ where the superconductivity of Ba$_{1-x}$K$_x$Fe$_2$As$_2$ emerges, it is very difficult to obtain highly homogeneous samples with a sharp single transition because of a very sensitive change in $T_c$ by a small variation in composition in this region (see the phase diagram in Ref.\onlinecite{avc12a} or in Fig. \ref{F3} below).   Therefore, in this synthetic procedure a closed metal tube, with minimal volume, is necessary to suppress vaporization of K/As at high temperatures. Also for better control of K/Ba ratio during the reaction, the annealing temperature was gradually increased from $800\,^{\circ}\mathrm{C}$ for $x = 0.9$ to $1100\,^{\circ}\mathrm{C}$ for  $x = 0.2$. About 10-20\% of additional KAs in the reaction mixture after the pre-reaction is needed to compensate the loss of K/As that occurs by the extended period of annealing at high temperature for all compositions. Handling of all materials was carried out under a dry argon atmosphere. 

 Low-field dc magnetization under pressure, was measured in a Quantum Design Magnetic Property Measurement System, MPMS-5, SQUID magnetometer using a  a commercial, HMD, Be-Cu piston-cylinder pressure cell. \cite{hmd}  Daphne oil 7373 was used as a pressure medium and superconducting Pb or Sn (to have its superconducting transition well separated from that of the sample) as a low-temperature pressure gauge. \cite{eil81a}.  The heat capacity was measured using a hybrid adiabatic relaxation technique of the heat capacity option in a Quantum Design Physical Property Measurement System, PPMS-9 or PPMS-14, instrument.

\section{Results} 
\subsection{Pressure dependence of $T_c$}

\begin{figure}
\begin{center}
\includegraphics[angle=0,width=120mm]{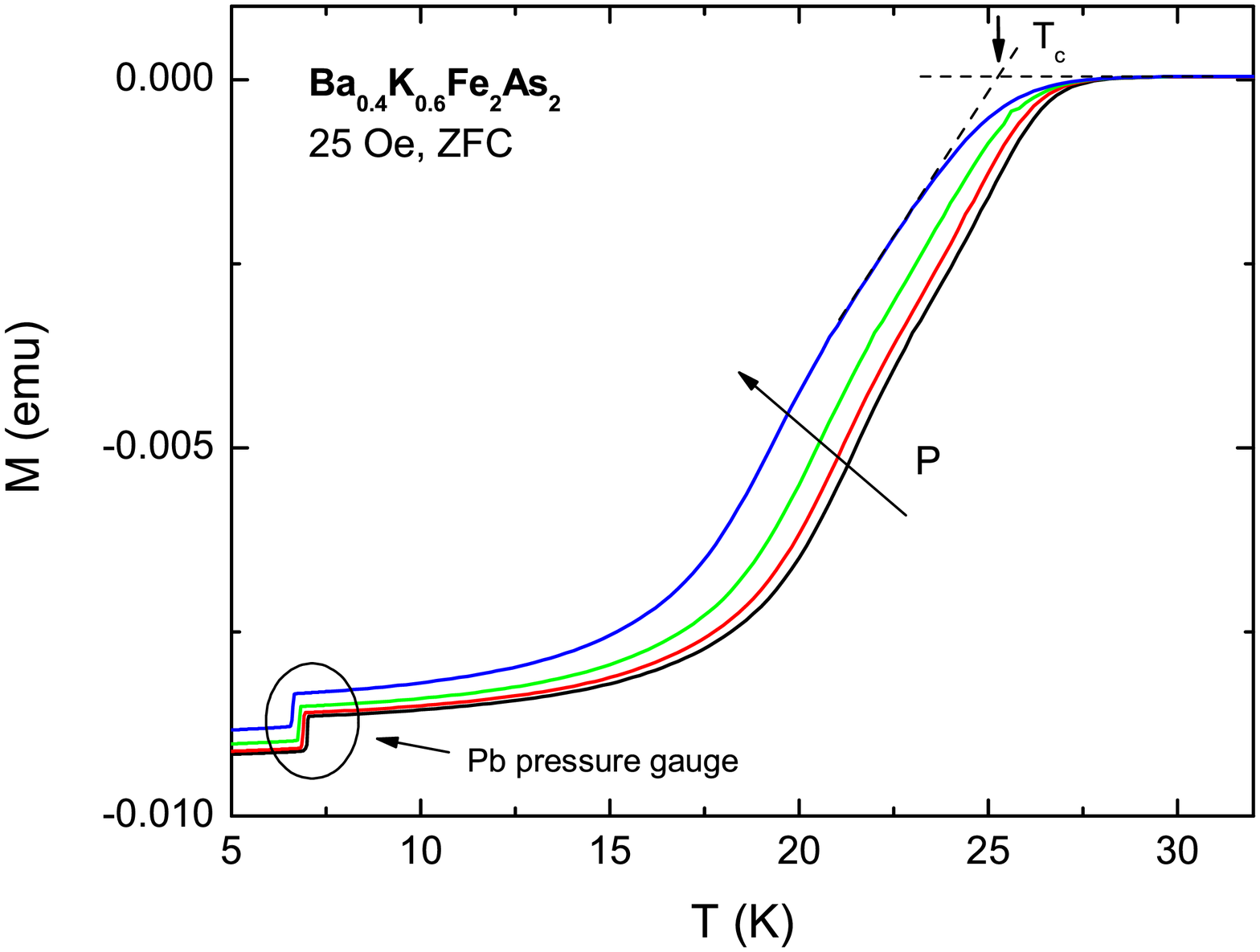}
\end{center}
\caption{(Color online)   Example of temperature dependent magnetization (zero-field-cooled, taken in 25 Oe applied magnetic field) of Ba$_{0.4}$K$_{0.6}$Fe$_2$As$_2$ measured at 1.5, 3.9, 6.5, and 11.1 kbar. The criterion for $T_c$ (for $P = 11.1$ kbar curve as an example) used in this work is shown by the arrow. Superconducting transitions  in Pb used as a pressure gauge are seen near 7 K.} \label{F1}
\end{figure}

\begin{figure}
\begin{center}
\includegraphics[angle=0,width=120mm]{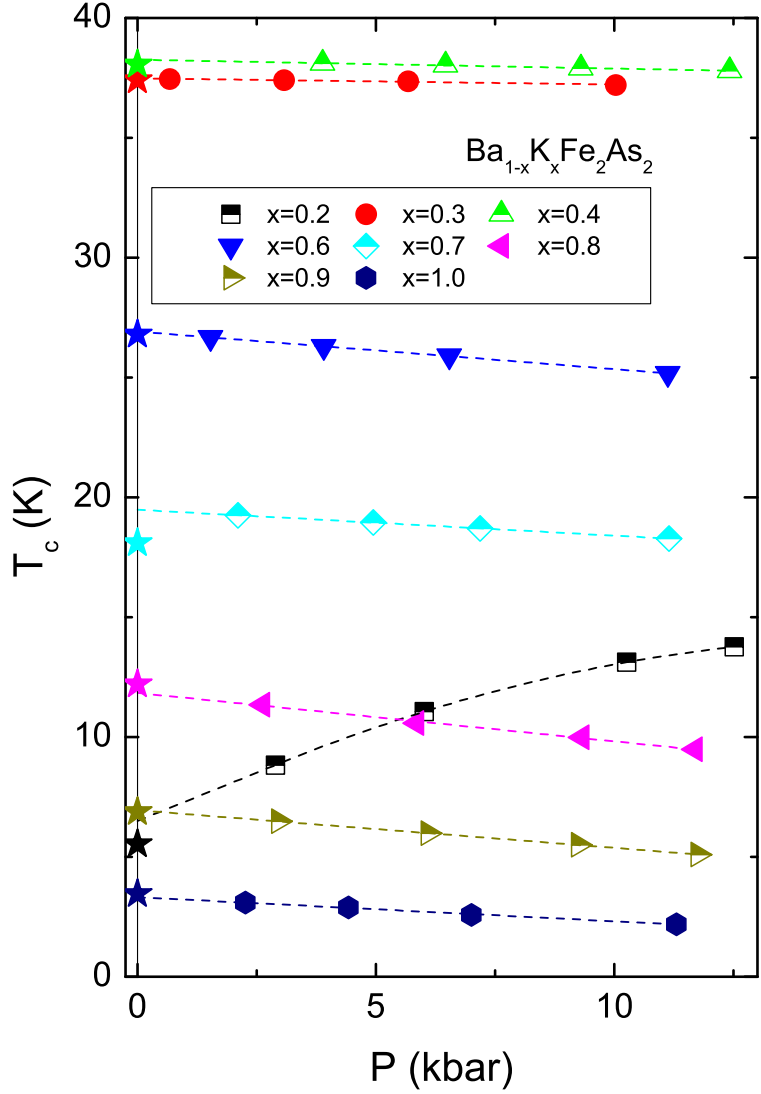}
\end{center}
\caption{(Color online) Summary plot of the pressure dependence of $T_c$ for the samples studied in this work. Dashed lines are linear fits to the data except for the $x = 0.2$ where the fit is a second order polynomial. These lines are extended to $P = 0$. Stars are the $T_c$ values obtained in the analysis of the ambient pressure heat capacity data below.} \label{F2}
\end{figure}

An example of $M(T)$ data taken at different pressures is shown in Fig. \ref{F1}. An onset criterion was used to determine $T_c$. Since none of the samples was suffering from a distinct change of the transition width under pressure, alternative criteria (like maximum in $dM/dT$) yield similar pressure dependencies. The pressure dependencies of the superconducting transition temperatures for the samples with all K-concentrations studied in this work are shown in Fig. \ref{F2}. For the underdoped sample, with $x = 0.2$, $T_c$ increases under pressure, for samples close to the optimally doped, $x = 0.3, 0.4$, the $T_c(P)$ dependencies are basically flat, and for the overdoped samples, $x \geq 0.6$, $T_c$ decreases under pressure.  For all samples, except the underdoped $x = 0.2$  the $T_c(P)$ behavior is linear, as such the pressure dependencies of $T_c$ can be well represented by $dT_c/dP$ values. The set of $T_c(P)$ data yields the ambient pressure $T_{c0}(x)$ values defined as extrapolation to $P = 0$ data under pressure. These values are consistent with the superconducting transition temperatures obtained from the analysis of the ambient pressure specific heat data below, shown as stars in Fig. \ref{F2}.

A more compact way to look at the pressure dependence of $T_c$ in the  Ba$_{1-x}$K$_x$Fe$_2$As$_2$ series is presented in Fig. \ref{F3}, where the results of this work are plotted together with the literature data \cite{bud12a,has12a} that show consistency between different groups/samples in the overlapping concentration regions. The $T_{c0}(x)$ data  are consistent with the ambient pressure data in the literature  \cite{rot08b, che09a,avc12a} and are shown for reference. The K-concentration dependent pressure derivatives, $d T_c/dP$, and normalized, $d(\ln T_c)/dP$ show a clear trend and three different K-concentrations regions. The pressure derivatives are positive and rather high for the significantly underdoped samples ($x \sim 0.2$, $T_c < 20$ K). They become small and negative for $0.3 \leq x \leq 0.7$, this concentration range covers slightly underdoped, optimally doped, and part of the overdoped samples; the concentration dependence of $d T_c/dP$ (or $d(\ln T_c)/dP$) is close to linear in this middle region. On further increase of $x$ the $x$-dependence of the pressure derivatives deviated from the shallow linear behavior (this trend is better seen in the concentration dependence of the normalized pressure derivatives).

\begin{figure}
\begin{center}
\includegraphics[angle=0,width=120mm]{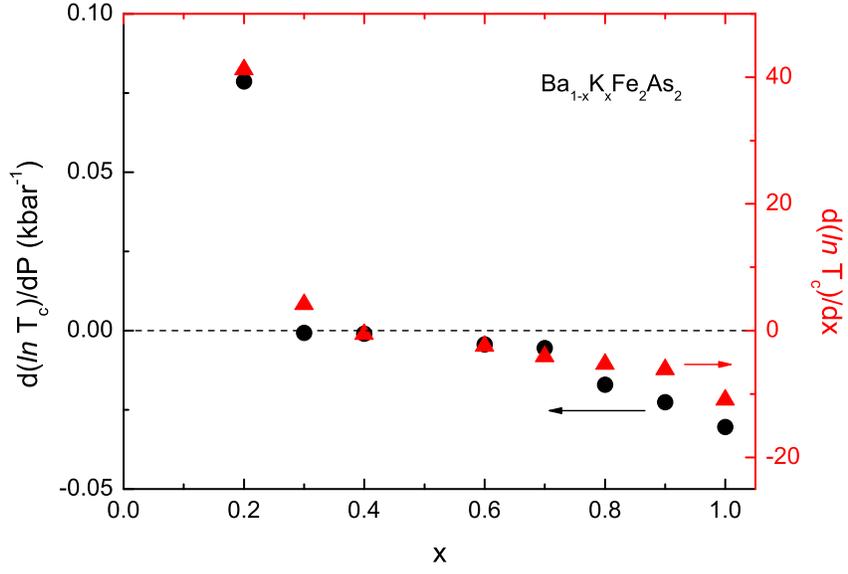}
\end{center}
\caption{(Color online) K - concentration dependence of the normalized pressure derivatives, $d(\ln T_c)/dP = \frac{1}{T_{c0}}~d T_c/dP$ (left axis, circles), and the normalized concentration derivatives, $d(\ln T_c)/dx = \frac{1}{T_{c0}}~d T_c/dx$ (right axis, triangles) of the superconducting transition temperatures.  Dashed line corresponds to the shared zero on $Y$-axes.} \label{F4}
\end{figure}

Fig. \ref{F4} presents a comparison of the relative changes in superconducting transition temperature under pressure and with K - doping.   For $0.2 \leq x \leq 0.7$ both sets of data can be scaled reasonably well, illustrating apparent equivalence of the effect of pressure and doping on $T_c$, suggested for other members of the 122 family. \cite{kim09a,dro10a,kli10a,kim11a} This scaling however fails for $0.7 < x \leq 1.0$.

\subsection{Jump in specific heat}

\begin{figure}
\begin{center}
\includegraphics[angle=0,width=120mm]{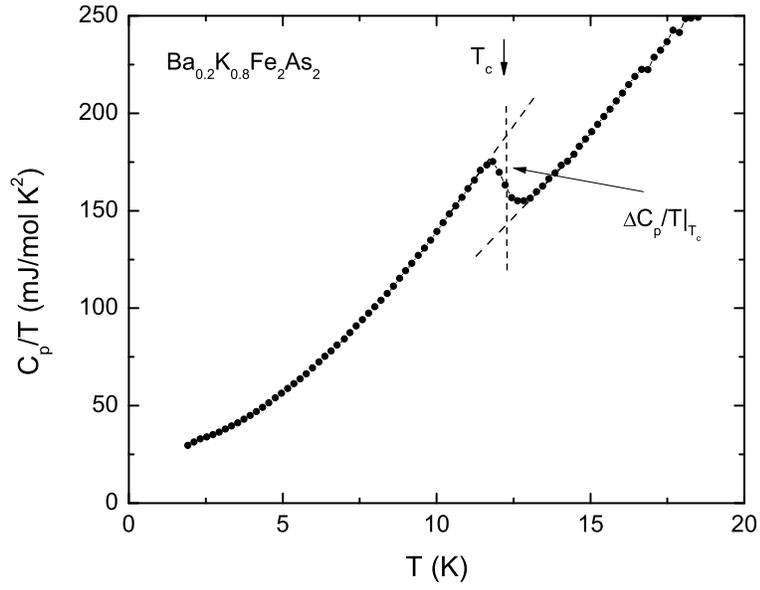}
\end{center}
\caption{Temperature-dependent heat capacity of  Ba$_{0.2}$K$_{0.8}$Fe$_2$As$_2$ plotted as $C_p/T$ vs $T$. Criteria for $T_c$ and  $\Delta C_p|_{T_c}$ (isoentropic construct) are shown.} \label{F5}
\end{figure}

\begin{figure}
\begin{center}
\includegraphics[angle=0,width=120mm]{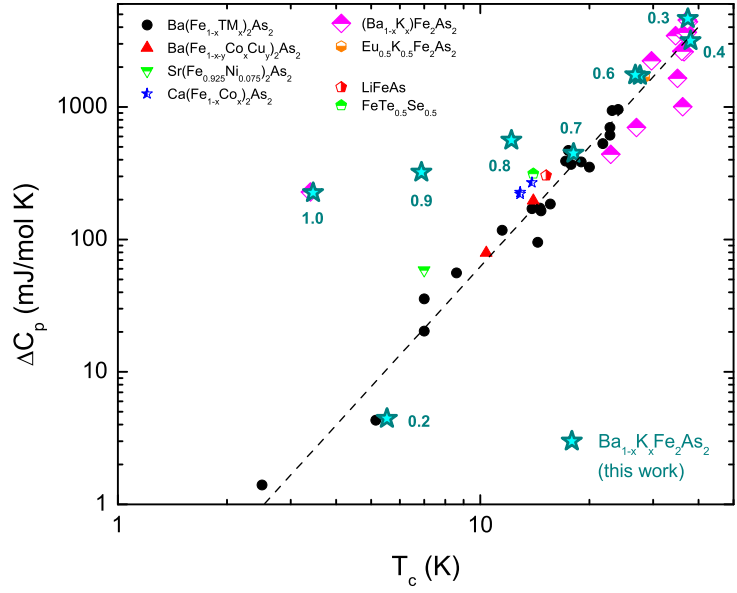}
\end{center}
\caption{(Color online) $\Delta C_p$ at the superconducting transition vs $T_c$  for the  Ba$_{1-x}$K$_x$Fe$_2$As$_2$ series, plotted together with literature data for various FeAs-based superconducting materials. Updated plot \cite{bud12a} is used to show the literature data. The line corresponds to $\Delta C_p \propto T_c^3$. Numbers near the symbols are K - concentration $x$. Two symbols for $x = 0.6$ correspond to measurements on two samples of slightly differing quality.} \label{F6}
\end{figure}

All samples studied in this work except for $x = 0.2$ show unambiguous jump in specific heat at $T_c$ (see Fig. \ref{F5} as an example). The $T_c$  and  $\Delta C_p|_{T_c}$ values were determined by a procedure consistent with that used in Ref. \onlinecite{bud09b}. For $x = 0.2$ the difference between the data taken in zero field and 140 kOe applied field was analyzed and the error bars in the obtained $\Delta C_p$ at $T_c$ value are expected to be rather large. The specific heat jump data for  the  Ba$_{1-x}$K$_x$Fe$_2$As$_2$ series obtained in this work were added in Fig. \ref{F6} to the updated BNC plot taken from Ref. \onlinecite{bud12a}. Again, there appears to be a clear trend: for $0.2 \leq x \leq 0.7$ the data follow the BNC scaling, in agreement with the scattered literature data for the samples with K - concentrations in this range. The data for $0.8 \leq x \leq 1.0$ clearly deviate from this scaling, with the data for the end-compound, KFe$_2$As$_2$, consistent with the previously published value. \cite{bud12a} This is clearly different from the studied Ba(Fe$_{1-x}${\it TM}$_x$)$_2$As$_2$ series, for which the BNC scaling was observed for the samples covering the full extent of the superconducting dome.

\section{Discussion and summary}

Both sets of measurements point out that some change in superconducting properties of Ba$_{1-x}$K$_x$Fe$_2$As$_2$ occurs near the $x = 0.7$  concentration range. NMR results \cite{hir12a} suggested that  the nodal line structure of the superconducting gap emerges at $x \sim 0.7$ without a change of the gap symmetry, although the possibility of the change  of the superconducting gap symmetry at this concentration was not completely excluded. These suggested changes were related to a Lifshitz transition, a disappearance of the electron pockets in the Fermi surface, at similar K - doping levels. \cite{hir12a} The anomaly in the evolution of the pressure derivatives with concentration is consistent with what one would expect in the case of the Lifshitz transition, \cite{mak65a} along the same lines as evolution of $dT_c/dP$ was understood in several cuprates. \cite{xio92a,cao95a} The observed violation of the BNC scaling hints that some significant changes might happen in the nature of the superconducting state as well. These results delineate the difference between transition metal doping  and K-doping in the BaFe$_2$As$_2$ series  concentration and define the K - concentration region where the superconducting state (possibly the order parameter) is significantly modified. Further measurements for concentrations around 70\% of K (in particular ARPES and thermal conductivity) would be desirable to understand the observed anomalies.

\begin{acknowledgments}

Work at the Ames Laboratory was supported by the US Department of Energy, Basic Energy Sciences, Division of Materials Sciences and Engineering under Contract No. DE-AC02-07CH11358. Work at the Argonne National Laboratory supported by the U.S. Department of Energy, Office of Basic Energy Sciences under contract No. DE-AC02-06CH11357. S.L.B. acknowledges partial support from the State of Iowa through Iowa State University.

\end{acknowledgments}

\clearpage

\clearpage

\clearpage

\begin{figure}
\begin{center}
\includegraphics[angle=0,width=120mm]{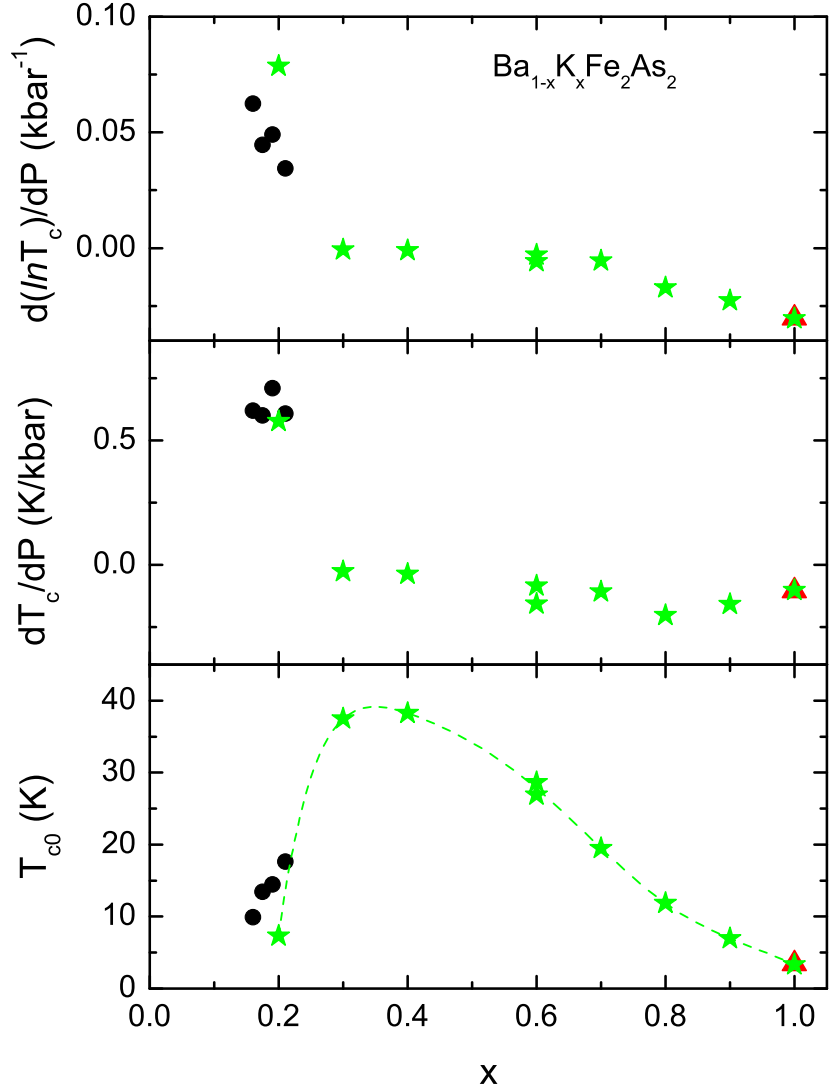}
\end{center}
\caption{(Color online) $T_c$, $d T_c/dP$ and $d(\ln T_c)/dP$ as a function of K-concentration in Ba$_{1-x}$K$_x$Fe$_2$As$_2$. Symbols: stars - this work, triangle - data for KFe$_2$As$_2$ from Ref. \onlinecite{bud12a}, circles - data for four underdoped samples obtained by linear fits of the first 3-4 data points under pressure in Ref. \onlinecite{has12a}. $T_{c0}$ values are taken from the linear fits of $T_c(P)$. Two symbols for $x = 0.6$ correspond to measurements on two samples of slightly differing quality. Dashed line is a guide for the eye.} \label{F3}
\end{figure}

\clearpage

\clearpage

\clearpage

\end{document}